\documentclass[review]{elsarticle}

\usepackage{lineno,hyperref}
\modulolinenumbers[5]

\journal{Comp Math Meth Med}

\usepackage{verbatim}
\usepackage{amsmath} 
\usepackage{graphicx}
\usepackage{grffile}
\usepackage{url}

\usepackage{etoolbox} 
\usepackage{breqn}   

\BeforeBeginEnvironment{dmath}{\par\begin{nolinenumbers}}%
\AfterEndEnvironment{dmath}{\end{nolinenumbers}}%
\BeforeBeginEnvironment{dmath*}{\par\begin{nolinenumbers}}%
\AfterEndEnvironment{dmath*}{\end{nolinenumbers}}
\BeforeBeginEnvironment{dgroup*}{\par\begin{nolinenumbers}}%
\AfterEndEnvironment{dgroup*}{\end{nolinenumbers}}

\newcommand\commentout[1]{}

\begin{document}

\begin{frontmatter}

\title{Hamiltonian analysis of subcritical stochastic epidemic dynamics}

\author[proctor]{Lee Worden\corref{corrauth}}
\cortext[corrauth]{Corresponding author}
\ead{worden.lee@gmail.com}

\author[nrl]{Ira B. Schwartz}

\author[ibm]{Simone Bianco}

\author[proctor,biostat]{Sarah~F.~Ackley}

\author[proctor,ophth]{Thomas~M.~Lietman}

\author[proctor,ophth,biostat]{Travis~C.~Porco}

\address[proctor]{Francis I. Proctor Foundation, University of California San Francisco, San Francisco, California, USA}
\address[nrl]{Nonlinear Systems Dynamics Section, Plasma Physics Division, U.S. Naval Research Laboratory, Washington, DC, USA}
\address[ibm]{Department of Industrial and Applied Genomics, IBM Accelerated Discovery Lab, IBM Almaden Research Center, 650 Harry Rd, San Jose, CA 95120-6099, USA}
\address[biostat]{Department of Epidemiology and Biostatistics, University of California San Francisco, San Francisco, California, USA}
\address[ophth]{Department of Epidemiology and Biostatistics, University of California, San Francisco, California, USA}

\begin{abstract}
We extend a technique of approximation of the long-term behavior of
a supercritical stochastic epidemic model, using the WKB approximation
and a Hamiltonian phase space, to the subcritical case.
The limiting behavior of the model and approximation are qualitatively
different in the subcritical case, requiring a novel analysis
of the limiting behavior of the Hamiltonian system away from its
deterministic subsystem.
This yields a novel, general technique of approximation of the quasistationary
distribution of stochastic epidemic and birth-death models,
and may lead to techniques for analysis of these models
beyond the quasistationary distribution.
For a classic SIS model, the approximation found for the quasistationary
distribution is very similar to published approximations but not identical.
For a birth-death process without depletion of susceptibles, the
approximation is exact.
Dynamics on the phase plane similar to those predicted by the Hamiltonian
analysis are demonstrated in cross-sectional data from trachoma treatment
trials in Ethiopia, in which declining prevalences are consistent
with subcritical epidemic dynamics.
\end{abstract}


\end{frontmatter}


\section{Introduction}

Stochastic models are a common tool in epidemiological
research, where public health
interventions aim at the reduction of fluctuating counts of
infected or infective individuals \cite{bailey1975}, and
models are used in explaining, predicting, and responding to
acute and chronic diseases of public health significance.

A fundamental result is the presence of a critical value of the basic
reproduction number $R_0$, defined as the expected number of
secondary cases resulting from a single infective case in an
otherwise susceptible population.
Supercritical diseases, those with $R_0>1$,
tend to stabilize around a positive number of infectives
that can persist for very long times,
while in subcritical cases ($R_0<1$) the infective count
declines to zero on a relatively short timescale.
In either case,
the long-term, stationary probability distribution of number of infectives
is trivial, as all epidemics
in finite-population stochastic transmission models 
must eventually die out due to chance fluctuations,
but the quasistationary distribution---the distribution
conditional on non-extinction of the disease---can be very 
informative about the behavior of the system within finite time
intervals.

When $R_0<1$,
the quasistationary distribution of number infective
in simple transmission models
is often approximately geometric,
with probability of $I$ infectives proportional to $(R_0)^I$
\cite{nasell_quasi-stationary_1996,lambert2008population}.
Prevalences consistent with the geometric distribution,
when analyzed statistically across multiple locations simultaneously,
have been observed in trachoma elimination trials
at times in which the disease's dynamics are subcritical
\cite{lietman_epidemiological_2011,lietman-gebre-abdou2015,rahman_distribution_2015}.

Such statistics of case count distributions
observed in multiple communities at a single time
may be able to help provide an assessment of the dynamics
of a disease, possibly of its basic reproductive number,
and hence, of the future time course of the disease.
An approximately geometric distribution of prevalences also implies
that there will be more high-prevalence communities than there would be
in a lighter-tailed distribution,
even when the mean prevalence is low and declining.
This suggests that an exceptionally high-prevalence community may be
simply a statistical outlier, which can be expected to regress to the
mean without intervention, rather than a ``transmission hotspot''
calling for intensified intervention \cite{lietman-gebre-abdou2015}.

While the quasistationary distribution of a specific stochastic model
can be calculated as an eigenvector of a Markov transition matrix,
since the equations for the entries of that vector
can not be solved explicitly for even very simple models,
research has focused on approximations
\cite[for example]{cavender_quasi-stationary_1978,kryscio_extinction_1989,nasell_quasi-stationary_1996,naasell2003moment}.
Barbour and Pollett \cite{barbour2010total} established that the
quasistationary distribution is a fixed point of a given map defined
on probability mass functions, allowing efficient approximation techniques
\cite{van2013quasi}. The fixed point of that map can also be
found using a ``ratio of means'' approach built on waiting times rather than
transition rates \cite{artalejo_quasi-stationary_2010}
that can aid in calculation.
Quasistationary approximations for diffusion processes
and branching processes are also well developed and
are the subject of active research and development
\cite{van2013quasi,lambert2008population,meleard2012quasi}.

In this paper we introduce a method of approximating the
quasistationary distribution of a stochastic model in the subcritical regime,
using a technique that has been used previously 
to approximate rare large-deviation events in supercritical dynamics
\cite{dykman_disease_2008,ovaskainen_stochastic_2010,schwartz_converging_2011}. 
This technique takes a large-population limit of the model
dynamics in a way that yields a Hamilton-Jacobi equation,
which can be understood by analyzing the geometry of an
associated Hamiltonian ODE system.

This Hamiltonian approach to stochastic mechanics, innovated by Graham
\cite{graham_weak-noise_1984} for diffusion equations
and extended by Hu \cite{hu_stationary_1987} to master equations,
has primarily been used to study stationary solutions of the limiting
stochastic process, by locating special solutions of the Hamiltonian
ODE system, characterized by $H=0$ where $H$ is the Hamiltonian.
The Hamiltonian ODE system includes the
deterministic limit of the stochastic model as an invariant subsystem
within the equipotential ($H=0$) set,
and at each limit set of the deterministic system,
the equipotential set extends outwards into the non-deterministic
regions of the Hamiltonian system's phase space.
Those extensions reveal quantitative information about
the system's stochastic behavior near attractors.
Thus they are used to analyze stationary probability densities associated
with attractors and other limit sets of the deterministic system,
and the frequencies and paths of rare escape events from one attractor
to another
\cite{ovaskainen_stochastic_2010,black_wkb_2011,forgoston2011maximal,lindley_iterative_2013,lindley_rare-event_2014}.
This geometric structure, which encodes characteristics of the deterministic
limit of the stochastic system and the probability distribution of
deviations from the deterministic limit, is strange in comparison to
the structures seen in Hamiltonian systems from physics, and is
much less well understood.

Here we investigate the use of
structures within the equipotential set, but at a distance
from the deterministic subsystem, to analyze a stochastic model's behavior.
We identify such a structure far from the deterministic subsystem
with the quasistationary behavior of an epidemic model,
in contrast to the use of structures intersecting the deterministic subsystem
to analyze stationary behavior.

\section{Limiting behavior of birth-death process}
\label{sec:bd}

Many models of stochastic epidemic dynamics, biological population dynamics
more generally, and branching processes, are included in the category of
birth-death processes. Here we apply the analysis of Hu
\cite{hu_stationary_1987} to
this class of processes, and below we will apply it to specific example
models.

A stochastic birth-death process models the size of a single population,
altered by events in which the size either increases by one or decreases
by one. The rate of increase from size $k$ is labeled $B(k)$ and
the rate of decrease from size $k$ is labeled $D(k)$.
Writing $P(k,t)$ for the probability that the size is $k$ at time $t$,
the change in probability over time
is governed by a master equation:
\begin{dmath} \label{eqn:BD-master}
\frac{dP(k,t)}{dt} = B(k-1)P(k-1,t) + D(k+1)P(k+1,t) - B(k)P(k,t) - D(k)P(k,t)
\quad\text{for each }k.
\end{dmath}
Taking $D(0)=0$ and $B(-1)P(-1,t)=0$ for all $t$,
the dynamics of the master equation is confined to nonnegative
values of $k$.
In order to take a large-system-size limit, let 
$\Omega$ be a measure of system size
such as, for example, a maximum population size,
such that as we consider increasingly large
birth-death systems in which both $\Omega$ and $k$
become unboundedly large,
the ratio $k/\Omega$ remains finite.
For example, in a system with finite population size $N$, we can
use $\Omega=N$, as we will see below.
Then letting $x=k/\Omega$, we obtain a transformed master equation
\begin{dmath*}
\frac1\Omega \frac{dP(x,t)}{dt} =
  b\left(x-\frac1\Omega\right) P\left(x-\frac1\Omega\right,t)
  + d\left(x+\frac1\Omega\right) P\left(x+\frac1\Omega\right,t)
  - b(x)P(x,t) - d(x)P(x,t),
\end{dmath*}
where $b(x)=(1/\Omega)B(\Omega x)$ and $d(x)=(1/\Omega)D(\Omega x)$.
Let the functions $b$ and $d$ be smooth functions of $x$ for each $\Omega$,
with a smooth limit as $\Omega\to\infty$.

Additionally, let $\phi(x,t)$ be
a probability density function
that is smooth in $x$ and $t$, such that
$\phi(k/\Omega,t)=\Omega\,P(k/\Omega,t)$.
Following Hu \cite{hu_stationary_1987},
this allows construction of a Kramers-Moyal expansion
of the dynamics, by substituting and
Taylor expanding the master equation around $x$
so that it is expressed using only values at $x$:
\begin{dmath}[label=BD-KM]
 \frac1\Omega \frac{\partial \phi(x,t)}{\partial t} = \sum_{n=1}^\infty \frac1{n!} \left( - \frac1\Omega \right)^n \frac{\partial^n}{\partial x^n} \left( b(x) \phi(x,t) \right) + \sum_{n=1}^\infty \frac1{n!} \left( \frac1\Omega \right)^n \frac{\partial^n}{\partial x^n} \left( d(x) \phi(x,t) \right).
\end{dmath}

To derive a partial differential equation in the large-system limit,
we rewrite the density as an exponential expression:
\begin{dmath}[label=fs]
\phi(x,t) = \Omega e^{-\Omega U(x,t)}.
\end{dmath}
Assume that the function $U$ can be expanded in powers of $\Omega$
on $0<x<1$:
\begin{dmath*}
U(x,t) = u(x,t) + \frac1\Omega u_1(x,t) + \frac{1}{\Omega^2}u_2(x,t) + \cdots,
\end{dmath*}
and that the terms of that expansion
other than $u(x,t)$ vanish asymptotically as
$\Omega$ approaches infinity.
This \emph{ansatz}, known as the WKB approximation
\cite{hu_stationary_1987,bender_orzag_1978},
makes it possible to generate a partial differential equation in $u$.

With these assumptions,
derivatives of products of $\phi$ take on a simplified form,
\begin{dmath*}
\left[-\frac{1}{\Omega}\right]^n\frac{\partial^n}{\partial x^n}F(x,t)e^{-\Omega U(x,t)} =
  e^{-\Omega U(x,t)}F(x,t)\left(\frac{\partial u}{\partial x}\right)^n + \mathcal{O}\left(\frac1\Omega\right).
\end{dmath*}
Substituting, the expansion of (\ref{BD-KM}) to first order is
\begin{dmath*}
\frac1\Omega \frac{\partial \phi(x,t)}{\partial t}
 = \Omega \left[ e^{-\Omega U(x,t)} \left( b(x) \sum_{n=1}^\infty \frac1{n!} \left(\frac{\partial u}{\partial x}\right)^n + d(x) \sum_{n=1}^\infty \frac1{n!} \left(-\frac{\partial u}{\partial x}\right)^n \right) + \mathcal{O}\left(\frac1\Omega\right) \right].
\end{dmath*}
Thus, in the large-size limit,
(\ref{BD-KM}) becomes
a partial differential equation for $u$:
\begin{dmath}[label=BD-HJ]
\frac{\partial u(x,t)}{\partial t} = - \left( b(x) \left(e^{{\partial u}/{\partial x}}-1\right) + d(x) \left(e^{-{\partial u}/{\partial x}}-1\right) \right).
\end{dmath}

\subsection{The associated Hamiltonian system}

Because the right hand side of (\ref{BD-HJ}) contains only first partial
derivatives of $u$,
it has the form of a
Hamilton-Jacobi equation of classical mechanics \cite{courant_methods_1989},
\begin{dmath*} \frac{\partial u(x,t)}{\partial t} = - H\left(x,\frac{\partial u}{\partial x}\right),
\end{dmath*}
with the consequence that it can be analyzed using characteristic
curves described by an associated system of ordinary differential equations
\cite{hu_stationary_1987}.
This analysis is based on the Hamiltonian function
\begin{dmath*} H\left(x,\frac{\partial u}{\partial x}\right) = b(x) \left(e^{{\partial u}/{\partial x}}-1\right) + d(x) \left(e^{-{\partial u}/{\partial x}}-1\right).
\end{dmath*}

From that Hamiltonian can be written a two-dimensional dynamical
system, whose state variables are $x$, the scaled population size, and a
conjugate variable $p$, which takes the place of
${\partial u}/{\partial x}$ in the Hamiltonian.  The associated
Hamiltonian dynamical system is
\begin{dmath}[label=BD-Hamiltonian-dynamics]
\begin{array}{r@{}r@{}l}
\dfrac{dx}{dt} ={}& \dfrac{\partial}{\partial p}H(x,p)   &{}= b(x) e^p - d(x) e^{-p} \\[12pt]
\dfrac{dp}{dt} ={}& - \dfrac{\partial}{\partial x}H(x,p) &{}= - b'(x) \left(e^p-1\right) - d'(x) \left(e^{-p}-1\right).
\end{array}
\end{dmath}
Trajectories of this system do not correspond to realizations of the
stochastic birth-death process,
but rather trace out curves along the surface of $u$
versus $x$ and $t$, which can be used to analyze the behavior
of $u$ over time.

Thus we can gain information about birth-death processes
in the large size limit
by using this associated system to analyze the
Hamilton-Jacobi equation (\ref{BD-HJ}).
Stationary solutions of the master equation,
characterized by the equilibrium condition $d\phi(x,t)/dt=0$,
are identified with curves on the $(x,p)$ plane on which $H(x,p)=0$.

In the case of this one-dimensional system, though not in the general
master-equation case, the Hamiltonian has two factors,
\begin{dmath}\label{eqn:BD-H}
 H(x,p) = \left( b(x) - d(x)\,e^{-p} \right) \left(e^p - 1\right),
\end{dmath}
which contribute two solution sets to the solution of $H=0$.

The flat subspace $p=0$ is always a solution set for $H=0$
in Hamiltonian systems constructed from master equations in this way
\cite{hu_stationary_1987}.
The dynamics within this set are the dynamics of the ODE
approximation to the stochastic dynamics, and fixed points
and other limit sets
of the Hamiltonian system located in this set correspond
to fixed points and other limit sets of this deterministic subsystem.
Other solutions to the equation $H=0$ pass transversely through
those limit sets, and can reveal information about the
stochastic behavior of the master equation system, as we will
see in the treatment of the supercritical SIS model, below.

In the birth-death systems we consider here, in which $k=0$ is
an absorbing state, a common factor of $x$ can be taken out of
$b(x)$ and $d(x)$,
allowing us to describe three components
of the solution set in all.

\section{The SIS model}

The SIS (susceptible-infective-susceptible) model
provides a simple representation of
infectious disease processes in the absence of immunity \cite{hethcote1976}.  
Classically, this model describes the number of susceptibles $S$ and infectives $I$
in a population of fixed size, where increase in the infective class
is driven by infective-susceptible contact events, and infectives
return to the susceptible class at a rate independent of contact with others.
SIS models have been used to describe a range of diseases, including
trachoma
\cite{lietman-porco-dawson1999,liu-porco-amza2015b,liu-porco-amza2015,liu-porco-mkocha2014,liu-porco-ray2013}
and sexually transmitted infections \cite{lajmanovich-yorke1976,hethcote-yorke1984}.
In population biology, a model identical in form to this one is known
as a stochastic logistic model \cite{nasell_extinction_2011}.

In the basic SIS model, the infective class increases at a rate
$\beta S (I / N)$, which is proportional to a quadratic susceptible-infective contact rate,
and decreases at a per capita constant rate $\gamma I$,
with $S=N-I$, and total population $N$ held fixed. 
Thus it is the number infective, $I$, that
is the stochastically varying state variable of the model.
Infective cases are added by transmission events, at rate
$\beta\,(S/N)\,I$, where
$\beta$ is the transmission rate per susceptible-infective pair
\cite{bailey1975}.
Cases return to the susceptible class at rate $\gamma\,I$,
where $\gamma$ is the per capita removal rate.
The parameters can be combined into one nondimensional value by
rescaling the time variable by a factor of $\gamma$,
after which the birth and death rates are
\[  B(I) = R_0 \left(1-\frac{I}N\right) I, \qquad D(I) = I,  \]
where $R_0=\beta/\gamma$ is the basic reproduction number
\cite{hethcote2000mathematics}.

\commentout{
The primary line of inquiry into the quasistationary behavior
of this model begins with Cavender's \cite{cavender_quasi-stationary_1978}
construction of the stationary distribution of a closely-related process,
a modification of the SIS process in which the rate of transition from one
infective individual to none is set to zero.  The stationary distribution
$\mathbf{p}^{(0)}$ of this process is used as an approximation to the
quasistationary distribution of the SIS model. That approximation is
commonly studied together with the one introduced by Kryscio and
Lef\`evre \cite{kryscio_extinction_1989}, which takes the stationary
distribution $\mathbf{p}^{(1)}$ of the SIS process
modified by introducing one permanently
infective individual as an approximation to $\mathbf{q}$.
More recently, N{\aa}sell has defined a ``uniform approximation'', constructed
by applying an iterated contracting map to the starting vectors
$\mathbf{p}^{(0)}$ and $\mathbf{p}^{(1)}$, which approximates the true
quasistationary solution quite well in the body and left tail of the
distribution \cite{nasell_extinction_2007,nasell_extinction_2011}.
Many of the other approaches discussed in our introduction have also
been applied to the SIS model.
}

Using system size $\Omega=N$, the analysis we have presented for
birth-death systems applies to the SIS model, with Hamiltonian 
\begin{dmath*}
  H(x,p) = R_0 (1-x) x \left( e^p - 1 \right) - x \left( e^{-p} - 1 \right),
\end{dmath*}
where $x=I/N$ is the infective fraction of the population.

\subsection{The supercritical case}

In the supercritical ($R_0>1$) case, the SIS process is attracted to
a positive, or endemic, equilibrium value $x=1-1/R_0$, at which the birth and
death rates are equal.
The probability density of the fraction infective concentrates around
that value.
On very long time scales, however, in finite systems,
stochastic fluctuation will bring the
fraction infective to zero, which is an absorbing state from which
the epidemic cannot return.
Thus the stationary distribution of the process is a point mass at
$x=0$,
and the density function concentrated around the endemic equilibrium,
while it is a stationary distribution in the infinite-size limit,
is the quasistationary distribution in the finite cases.

\begin{figure}
\centering
\includegraphics[width=0.5\textwidth]{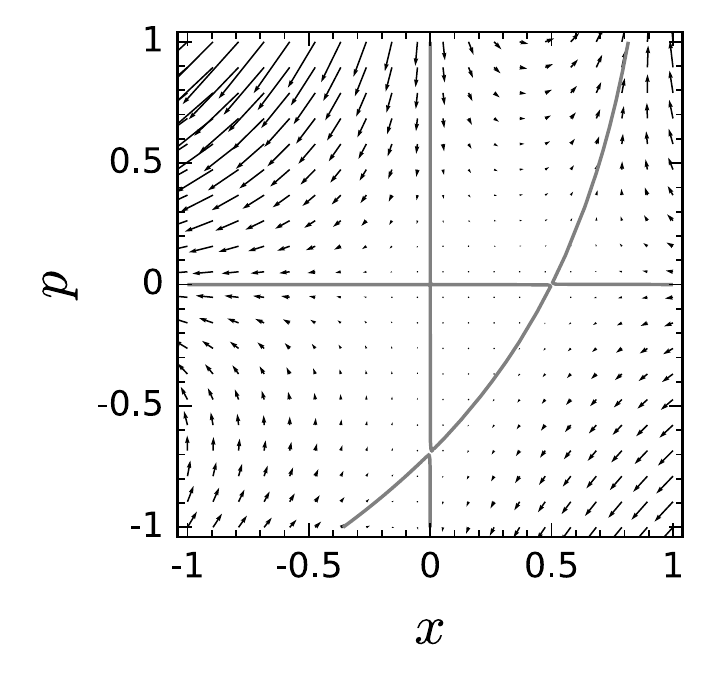}
\caption{\label{fig:super-plane}
{\bf Phase plane of the Hamiltonian dynamical system (\ref{BD-Hamiltonian-dynamics}),
for a supercritical SIS model} ($R_0=2$).
Arrows depict the flow of the dynamics of $x$ and $p$.
The three invariant curves of the dynamics
(solution curves of $H=0$) are shown in gray:
the two axes of the space, and one nontrivial curve.
The nontrivial curve corresponds to the quasistationary
solution of the stochastic SIS model, as discussed in the text.
}
\end{figure}

The Hamiltonian analysis of the supercritical SIS model
has been treated exactly elsewhere
\cite{schwartz_converging_2011,forgoston2011maximal}.
The phase plane of the Hamiltonian system is
shown in figure~\ref{fig:super-plane}.

Stationary solutions of the PDE correspond to solutions of $H(x,p)=0$
on this plane, when $p$ is interpreted as
${\partial u}/{\partial x}$.
The Hamiltonian factors into three parts:
\begin{dmath*}
H(x,p) = x(R_0(1-x) - e^{-p})(e^p-1),
\end{dmath*}
which directly identifies the three solution curves
of $H=0$ in the plane:
two trivial solutions,
\begin{dgroup*}
\begin{dmath*}
x = 0,
\end{dmath*}\begin{dmath*}
p = 0,
\end{dmath*}
\end{dgroup*}
and one nontrivial solution,
\begin{dmath} \label{eqn:SIS-curve}
p = -\ln( R_0 (1-x) ),
\end{dmath}
shown in figure~\ref{fig:super-plane}.
These curves are trajectories of the Hamiltonian dynamical system
(\ref{BD-Hamiltonian-dynamics}).

The horizontal axis of the phase plane, which is the $p=0$ solution,
is isomorphic to
the deterministic SIS system.  Two of the fixed points of the Hamiltonian
system are the fixed points of that deterministic system --- the
disease-free equilibrium at $(0,0)$
and the endemic equilibrium at $(1-1/R_0,0)$.  They are located at the
points where the horizontal axis intersects the other two solution curves.
A third fixed point, at 
$(0,-\ln R_0)$, also corresponds to the disease-free state ($x=0$), but is at the
intersection of solution curves away from the horizontal axis. 

The nontrivial solution curve (\ref{eqn:SIS-curve})
corresponds to the stationary
solution of $u(x)$ on which probability concentrates around the endemic equilibrium,
and the fixed points on it describe
the probability density at the endemic and disease-free equilibria.
That solution is a function $u(x)$ that solves
\begin{dmath*}
\frac{\partial u(x)}{\partial x} = -\ln(R_0(1-x)).
\end{dmath*}
Changing variables to $s=1-x$ and integrating
produces a closed-form solution,
\begin{dmath*}
u(s) = s\ln(R_0 s) - s + C_0.
\end{dmath*}
This provides a closed-form solution for the quasistationary probability
density:
\begin{dmath}[label=eq:approx]
\phi(s) = N e^{-Nu(s)}\ \hiderel{=}\ C_1\left(\frac{e}{R_0 s}\right)^{Ns}.
\end{dmath}
The constant $C_1$ is determined by the constraint that
$\int_0^1 \phi(s) ds = 1$.

In supercritical models in general, the equipotential surfaces
(solutions of $H=0$)
near the nontrivial solution of the deterministic
subsystem 
describe the behavior of the probability distribution of rare
events, which are located in the tail of the stationary distribution.

The above stationary solution approximates the quasistationary density
in the finite-$N$ SIS system,
in which extinction is a rare event given large $N$.

It provides an approximation for the time to extinction in the stochastic dynamics.
The function $u$ is the {\it action} of classical mechanics.
The most probable path to extinction can be obtained by maximizing the
function $u(x)$, which produces the equipotential surfaces $H=0$.
The path is explicitly calculated by integrating along the $H=0$ curves,
both in this SIS case and in more complex models 
(\emph{e.g.}\ \cite{schwartz_converging_2011}).

\section{Subcritical dynamics}

In the deterministic SIS system in the subcritical case,
$x$ relaxes to zero for all initial conditions $0\leq x\leq1$.  
The master equation solution also relaxes to $x=0$,
with probability mass declining to zero at all other values of $x$
\cite{nasell_quasi-stationary_1996}.
In this case, the quasistationary distribution is not stationary even
in the large-$N$ limit
due to the deterministic attraction of the origin.
The WKB hypothesis that the probability current near the
absorbing state $x=0$ vanishes when the system size $N$ grows
without bound is not satisfied, and we do not use the stationary
behavior of the PDE (which relaxes to a point mass) to analyze
the quasistationary behavior of the master equations.
Instead we use the transient behavior of the PDE to identify
the equilibrium structure in the Hamiltonian phase plane that
describes the master equation's quasistationary solution.

\subsection{Using the phase plane to analyze dynamics of the Hamilton-Jacobi equation}

\begin{figure}
\centering
\includegraphics[height=0.5\textwidth]{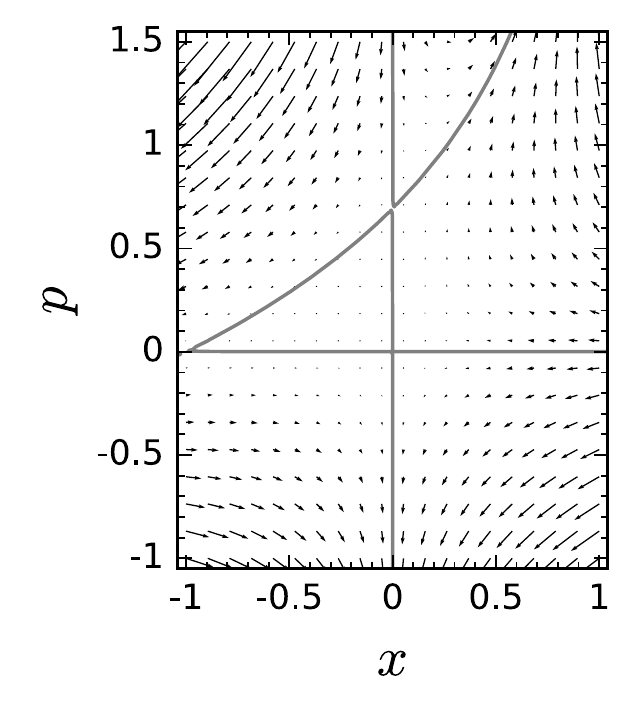}
\caption{ \label{fig:sub-plane}
{\bf Phase plane of Hamiltonian dynamical system for subcritical SIS system} ($R_0=0.5$).
Flow is represented by arrows and the three invariant curves of the dynamics
(solution curves of $H=0$) are shown in gray, as in Figure~\ref{fig:super-plane}.
In this case, the nontrivial curve is shifted to a different position,
and its intersections with the axes are located above and to the
left of the origin, where in the supercritical case (Figure~\ref{fig:super-plane})
they are below and to the right of the origin.
This leads to qualitatively different dynamics,
requiring a different analysis to explain the quasistationary
behavior of the model.
}
\end{figure}

In the Hamiltonian phase plane for the subcritical model,
the same three solution curves for $H=0$ are present as in the
supercritical case, but
they fall in different places on the phase plane, as shown in
figure~\ref{fig:sub-plane}.
In this case, the point of intersection of
the nontrivial curve (\ref{eqn:SIS-curve}) and the horizontal axis
is shifted to the left of the origin.
The endemic equilibrium represented by that point is lost
in a transcritical bifurcation when
$R_0$ declines below 1, and the origin becomes the attracting
solution for the stochastic SIS system.
The intercept where the nontrivial curve (\ref{eqn:SIS-curve})
meets the vertical axis,
at $p=-\ln R_0$, is now above $p=0$. 

Because of this bifurcation,
in the subcritical case
we cannot apply the analysis used for the supercritical case,
as the system is drawn to a singular value of $x$ at which the
$H=0$ curve crossing the horizontal axis is vertical, and can not be
translated to values of ${\partial u}/{\partial x}$ as a function
of $x$.
To study the quasistationary distribution of this system
requires further analysis.

\begin{figure}
\centering
\includegraphics[height=0.5\textwidth]{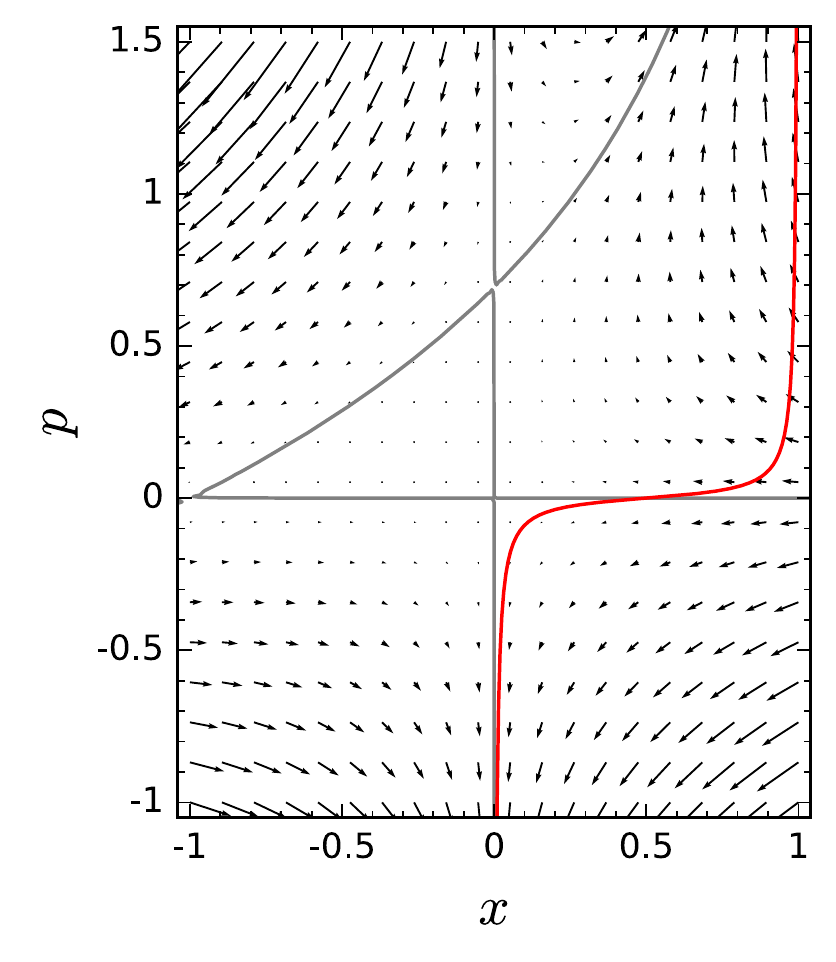}
\caption{ \label{fig:initial-u}
{\bf Initial condition for the subcritical SIS system} on
the Hamiltonian phase plane,
represented by a curve of $p$ values as a function of $x$.
In this and following figures, the initial condition used is
a $\beta$ distribution with $\alpha=\beta=2$, i.e.
$\phi_0(x)=6x(1-x)$, and using $N=100$,
transformed to a curve in the $x$-$p$ plane
using the relations $u(x)=-\ln({\phi_0(x)}/N)/N$
and $p=\partial u/\partial x$.
}
\end{figure}

Any smooth initial distribution $\phi(x)$ can be mapped onto a curve
in the $(x,p)$ plane
on which $p={\partial u}/{\partial x}$ at every value of $x$,
where $u$ is defined by $\phi(x) = Ne^{-Nu(x)}$ as above.
This curve for an example initial distribution is
plotted in figure~\ref{fig:initial-u}.

Integrating points of this curve
forward along trajectories of this system produces a geometric
representation of the time
evolution of the system as a moving curve in the phase plane,
on which the changing shape of ${\partial u}/{\partial x}$
is visible, and 
that relation between $\partial u/\partial x$ and $x$
provides information about the form of the function $u(x)$.

In terms of Hamiltonian dynamics, the function $u(x,t)$
is the \emph{action} of the system,
a scalar quantity that can be evaluated by integrating along its trajectories:
\begin{dmath*}
\frac{du(x,t)}{dt} = \frac{\partial u}{\partial x}\frac{dx}{dt} + \frac{\partial u}{\partial t}
 = p \frac{\partial H}{\partial p} - H.
\end{dmath*}

For convenience, it is possible to calculate $u$ directly when integrating
the Hamiltonian dynamics numerically,
by extending the dynamical system to include
$u$ as a state variable:
\newcommand*{\dhdpstrut}{\vphantom{\displaystyle\frac{\partial H}{\partial p}}}
\begin{dmath*}
 \frac{\partial}{\partial t} \begin{pmatrix}
	x \dhdpstrut\\[12pt]
	p \dhdpstrut\\[12pt]
	u \dhdpstrut \end{pmatrix}
   = \begin{pmatrix}
	\displaystyle \frac{\partial H}{\partial p} \\[12pt]
        \displaystyle -\frac{\partial H}{\partial x} \\[12pt]
        \displaystyle p\frac{\partial H}{\partial p} - H
     \end{pmatrix}.
\end{dmath*}
Assigning $u(x,0)=u_0(x)$ at each point of the initial curve and integrating
forward then yields values of $u(x,t)$ explicitly with increasing $t$.

\subsection{Evolution of the subcritical system from initial conditions}

\begin{figure}
\centering
\includegraphics[height=0.5\textwidth]{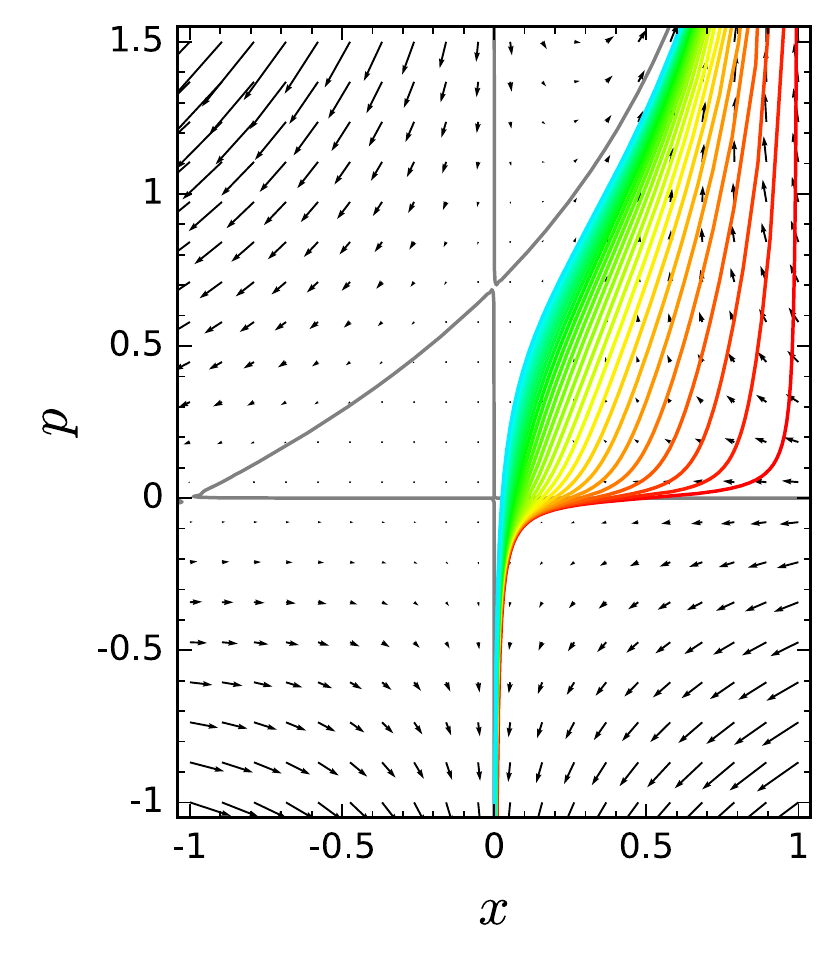}
\caption{ \label{fig:s-evolution}
{\bf Transient dynamics of the subcritical SIS system}
on the Hamiltonian phase plane,
evolving from the initial condition depicted in figure~\ref{fig:initial-u}
(red) toward later states (yellow, green, blue),
as each point of the initial curve moves according to the
Hamiltonian dynamics (\ref{BD-Hamiltonian-dynamics}).
}
\end{figure}

As time passes, each point of the $p$-versus-$x$ curve moves on the
phase plane according to the Hamiltonian dynamics.  Their evolution stretches
and translates the curve across the phase plane, as shown in
figure~\ref{fig:s-evolution}.  While any given point may move in somewhat
strange ways, including many that tend to infinity in the upper right
direction, the curve moves smoothly to the left,
approaching the vertical line $x=0$ and the gray curve that
extends into the first quadrant.

\begin{figure}
\centering
\includegraphics[width=0.5\textwidth]{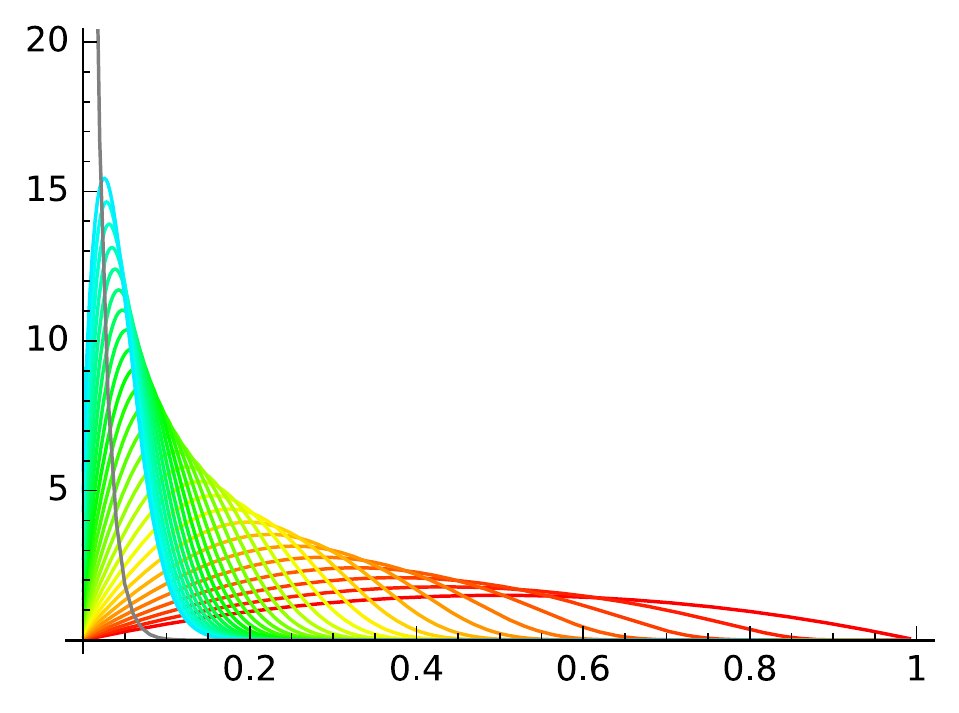}
\caption{\label{fig:phi-vs-t}
{\bf Transient dynamics of probability density}
in the subcritical SIS system,
displayed as $\phi(x,t) = Ce^{-Nu(x,t)}$ versus $x$
using the same data points as in figure~\ref{fig:s-evolution},
with $N=100$.
Each curve is normalized to total probability one.
The quasistationary distribution (\ref{eqn:SIS-QS}) is plotted in gray.
}
\end{figure}

From the moving points $(x,p,u)$ of this curve,
a plot of $u$ versus $x$ can be constructed,
or of $\phi=Ne^{-Nu}$ versus $x$, at each time $t$.
Figure~\ref{fig:phi-vs-t} presents this plot of $\phi$ versus $x$
in time.
The peak of the probability density moves asymptotically toward $x=0$,
and there is a declining tail to the right of the peak.

A number of features of the evolution of $u(x,t)$ versus $x$ are visible in
this view of the dynamics.  As discussed above, the dynamics on the horizontal
axis of the phase plane
is identical to the usual deterministic ODE for the SIS system.
When $p$ is read as ${\partial u}/{\partial x}$, it follows that that
horizontal axis, where $p=0$, corresponds to the extrema of the potential function
$u(x,t)$ with respect to $x$.
In the case pictured in these figures,
the only extremum is a minimum of $u(x,t)$, which
is a maximum of $\phi(x,t)$.
This implies that the maximum point of
the probability density function $\phi$,
which is the mode of the probability distribution,
in the large-system approximation we are using (\ref{BD-HJ}),
moves in exact accordance with the deterministic SIS dynamics.

Regions of $x$ values for which a curve in the $x$-$p$ plane
is below the horizontal axis
are regions where ${\partial u}/{\partial x}<0$, and equivalently on which
$\phi(x,t)$ is increasing in $x$, and regions where the curve is above the axis
are where $\phi(x,t)$ is decreasing in $x$.
Near the vertical axis, the $p$-versus-$x$
curve diverges to $p=-\infty$.  The fact that $p$,
representing ${\partial u}/{\partial x}$,
becomes negatively infinite there strongly suggests that $u(x)$ is
divergent to $+\infty$ at $x=0$, and so that $\lim_{x\to0^+}\phi(x,t)=0$,
at least in cases
like the one illustrated in which $\phi(0)$ is zero in the initial
conditions.

If the Hamilton-Jacobi PDE (\ref{BD-HJ}) is used
to approximate any finite-$N$ system, by grouping the
probability density into bins of width $1/N$, the result will be that
probability mass accumulates in the bin that includes $x=0$, and all the
other bins contain a tail that is decreasing in $x$, and whose
total mass declines asymptotically to zero as $t\to\infty$.

Figure~\ref{fig:s-evolution} demonstrates that in the long term,
the $p$-versus-$x$ curve becomes asymptotically close to the union of the
vertical axis below the positive-$p$ equilibrium and the nontrivial
$H=0$ curve (\ref{eqn:SIS-curve}) at and above that equilibrium.
We conclude that as the probability density accumulates near $x=0$,
the shape of the tail of the density on $x>0$ approaches a function
described by the diagonal curve, which is the nontrivial solution
(\ref{eqn:SIS-curve}) of $H=0$.
That tail defines the conditional distribution of $x$ given $x>0$,
and therefore the limiting curve (\ref{eqn:SIS-curve})
should provide an approximation
for the quasistationary distribution of the SIS master equations.

\subsection{Explicit approximation for the quasistationary distribution}

From the above analysis we conclude that the quasistationary
probability density function
of the master equation system (\ref{eqn:BD-master})
is approximated by the density function represented by the nontrivial
$H=0$ curve (\ref{eqn:SIS-curve}).  This is solved in the
same way as in the supercritical case:
\begin{dmath}\label{eqn:SIS-QS}
  \phi(s) = C_1 \left(\frac{e}{R_0s}\right)^{Ns},
\end{dmath}
where $s=1-x$.

While in the supercritical case this density function has a mode
at the endemic value $s=1/R_0$,
in this case the density is greatest at $x=0$ ($s=1$), as the
function is monotonic decreasing on the interval $0<x<1$.

Changing variables back to the number infective, $I=Nx=N(1-s)$, 
the quasistationary approximation becomes
\begin{dmath} \label{eqn:SIS-QS-discrete}
  P(I) \hiderel{=} \frac{1}{N}\phi(1-I/N) = C_2 \left(\frac{eN}{R_0(N-I)}\right)^{N-I},
\end{dmath}
using the appropriate normalizing factor $C_2$ for this
discrete probability mass function.

This quasistationary approximation is closely related to the classical
approximation $p^{(1)}$ of Kryscio and Lef\`evre
\cite{kryscio_extinction_1989,nasell_quasi-stationary_1996}:%
\footnote{We are thankful to an anonymous reviewer for this
observation (and see also \cite{kurtz1971limit}).}
their approximation,
\begin{dmath*}
  p^{(1)}(I) = C_3 \frac{1}{(N-I)!} \left(\frac{R_0}{N}\right)^I,
\end{dmath*}
when transformed using Stirling's approximation for factorials,
\begin{dmath*}
  \ln n! \approx n\ln n - n,
\end{dmath*}
yields the approximation we have derived:
\begin{dmath*}
  p^{(1)}(I) 
    \approx C_3 \left(\frac{e}{N-I}\right)^{(N-I)} \left(\frac{N}{R_0}\right)^{-I}
    \approx C_4 \left(\frac{eN}{R_0(N-I)}\right)^{N-I}
\end{dmath*}
(where $C_3$, $C_4$ are normalizing constants).

Previous approximations and numeric evaluation have established
\cite{cavender_quasi-stationary_1978,kryscio_extinction_1989,nasell_quasi-stationary_1996}
that the quasistationary distribution of the subcritical SIS system
is approximately geometric near $I=0$, with the probabilities of successive
values of $I$ having ratio $R_0$.

Thus the approximating geometric distribution has the form
\begin{dmath*}
 \Gamma(I) = C_5 (R_0)^{I}.
\end{dmath*}
The geometric distribution is characterized by the constant slope
of its logarithm:
\begin{dmath*}
 \frac{d}{dI}\ln\Gamma(I) = \frac{d}{dI}\left[\ln C_5 + I\ln R_0\right]
   = \ln R_0.
\end{dmath*}

Comparing to our approximation $p$,
the slope of $\ln p$ is not constant:
\begin{dmath*}
 \frac{d}{dI}\ln P(I)
   = \frac{d}{dI}\left[ \ln C_2 + (N-I)\left( 1 + \ln N - \ln R_0 - \ln(N-I) \right) \right]
   = - \left( 1 + \ln N - \ln R_0 - \ln(N-I) \right) + (N-I) \left( \frac{1}{N-I} \right)
   = \ln R_0 + \ln\frac{N-I}{N}.
\end{dmath*}

However, near $I=0$, the non-constant term is approximately zero, and the slope
of the logarithm is approximately $\ln R_0$, with the consequence that the
distribution is approximately geometric with the desired ratio
when $I\ll N$.

Since the ratio $(N-I)/N$ is smaller than one when $0<I<N$,
and thus its logarithm is negative,
it follows
that the probability mass function $p$ decreases to zero more rapidly than
the geometric function $\Gamma$ does as $I$ increases.

In an appendix we compare the SIS process to a birth-death
process that has the transmission and removal rates
of the SIS model without the effect of depletion of susceptibles,
and whose quasistationary distribution
is exactly the geometric distribution that approximates the above
distribution. The phase plane analysis of the birth-death process
provides visual evidence that the parameter characterizing the
approximating geometric distribution by its rate of decay is
determined by the intercept where the nontrivial curve
(\ref{eqn:SIS-curve}) crosses the vertical axis.

\section{Statistics of declining trachoma case counts}

While the SIS model has proven a theoretically interesting,
simple model of disease transmission, as discussed above,
it has also been used in practice in trachoma research,
In trachoma research, it has been used to assess
to assess treatment frequency needed for elimination \cite{lietman-porco-dawson1999},
efficacy of antibiotic treatment \cite{liu-porco-mkocha2014},
and waning of immunity \cite{Liu2013}, and in forecasting \cite{liu-porco-amza2015},
among other applications
\cite{melese-chidambaram-alemayehu2004,Ray2007,Ray2009,Lietman2011,liu-porco-mkocha2014,liu-porco-amza2015b,Gao2016}.

Trachoma is a common subclinical childhood infection in
certain regions of the less-developed world.
Repeated infection results in scarring of the eyelid, trichiasis
(turning inward of the eyelashes, so that the eyelids scrape
against the cornea).
Millions of cases of blindness have resulted.
The causative agent, \emph{Chlamydia trachomatis},
can be cleared with high efficacy with a single dose of
azithromycin 
\cite{Schachter1999,Chidambaram2006}.
The World Health Organization currently recommends annual
mass treatment in affected communities as a public health
control measure
\cite{melese-chidambaram-alemayehu2004,solomon2006trachoma,Chidambaram2006,House2009}.

\begin{figure}
\centering
\includegraphics[width=0.5\textwidth]{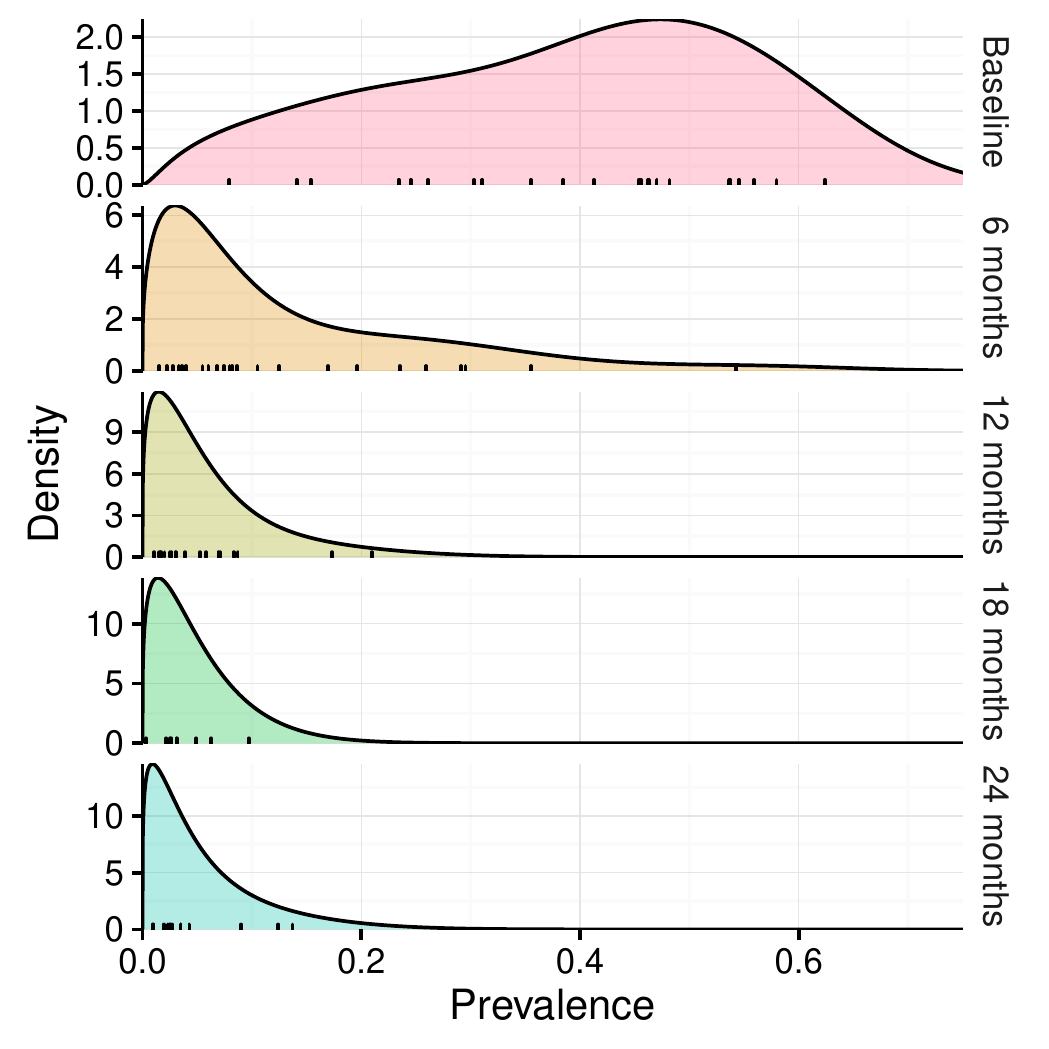}
\caption{ \label{fig:tana-density}
{\bf Changing trachoma prevalence} at baseline, and at 6-month intervals
during the TANA
trial of mass administration of azithromycin
\cite{Gebre2012}.
As the trial progresses, the prevalences become smaller
and become more closely approximated by the exponential 
\cite{lietman-gebre-abdou2015}.
(Individual village prevalences are shown in tick marks on the horizontal
axis. Curves result from beta distribution kernel density smoothing
\cite{chen1999beta},
with smoothing parameter determined from leave-one-out cross-validation
\cite{burnham1998model}.)
}
\end{figure}

During a clinical trial of timing of mass administration
of azithromycin in the Amhara Region of Ethiopia
\cite{House2009,Stoller2011,Gebre2012},
village-level prevalence data were collected.
At baseline the probability distribution of village-level prevalences,
omitting zero values, had a mean of 0.39 (range 0.08--0.62)
(figure~\ref{fig:tana-density}, top plot).
After the initiation of mass treatment at or exceeding recommended
WHO levels, the mean prevalence declined, and the distributions
became indistinguishable from exponential
\cite{lietman-gebre-abdou2015}
(figure~\ref{fig:tana-density}, subsequent plots).
This finding is consistent with the approximately exponential distributions
predicted by simple epidemic models, as discussed above.
The matter is of more than theoretical interest, as mentioned in our
introduction: the long tail of
the exponential distribution implies that during an elimination
campaign, some communities may have unexpectedly large prevalence
and appear to be outliers when in fact they are entirely consistent
with the variation expected.

\begin{figure}
\centering
\includegraphics[width=0.5\textwidth]{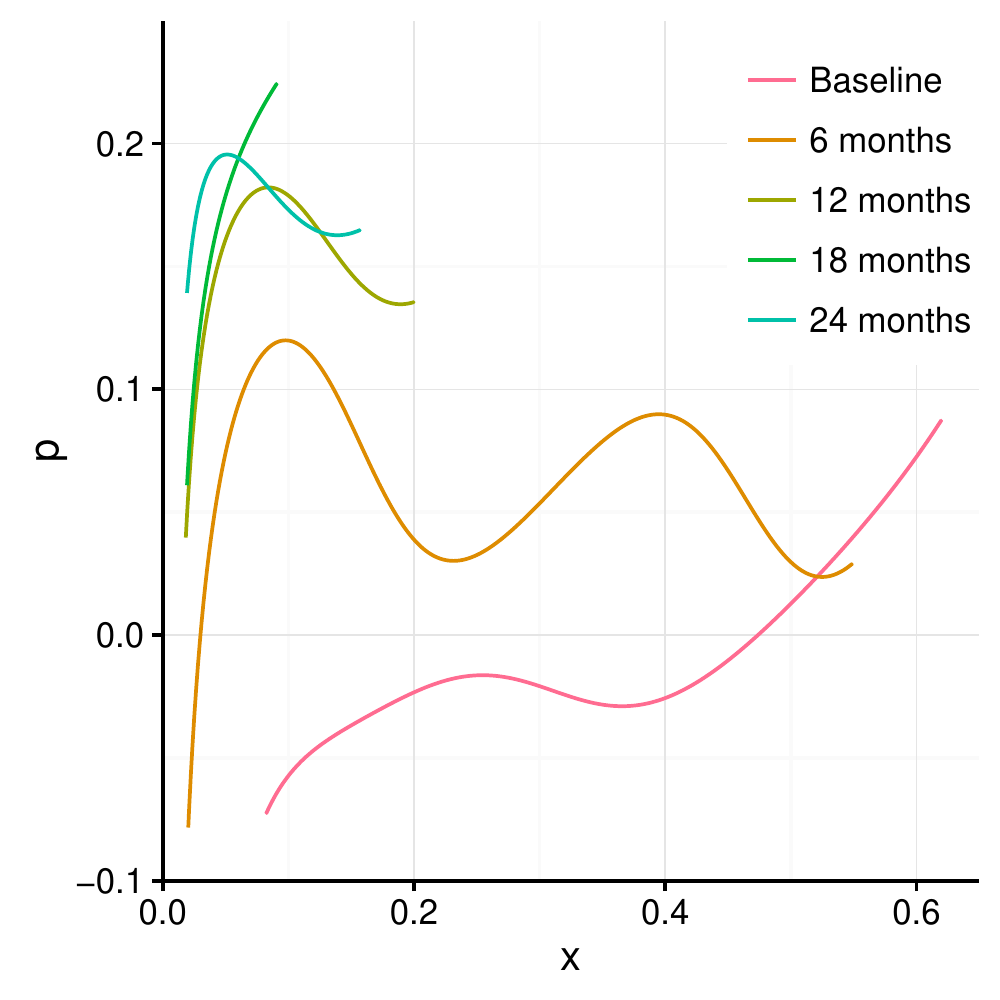}
\caption{ \label{fig:tana-phaseplane}
{\bf Phase plane representation of changing trachoma prevalence}
data from TANA trial shown in figure~\ref{fig:tana-density}.
Each curve on the plot corresponds to one of the distributions
shown in figure~\ref{fig:tana-density},
transformed to the $x$-$p$ plane as in earlier figures
(see text for details).
Over time the curves shift upward and to the left,
moving close to the vertical axis for smaller values of $p$
and diverging from it at larger values of $p$,
similar to the motion seen in the Hamiltonian analysis of
the SIS model (figure~\ref{fig:s-evolution}).
Each curve in this figure is restricted to the range of the nonzero
prevalence values.
}
\end{figure}

Figure~\ref{fig:tana-phaseplane} displays these probability
density functions $\phi(x)$
transformed to the phase-plane representation
defined above, $p(x) = -\frac{d}{dx}\ln(\phi(x)/N)/N$.
We assume a population size $N=100$ per village,
which is approximately the number of children at risk in
one of these villages
\cite{Gebre2012}.
In this plot, the same motion from lower right to upper left is
visible, with convergence to the vertical axis and possibly to
a curve leaving that axis in the positive quadrant.
More abundant data may permit location of such
a limiting curve that would intersect the vertical axis in this
representation of the data.
That curve would provide an estimate of the quasistationary behavior of
the disease, and its intercept would provide an estimate of the
disease's $R_0$.

\section{Summary}

Hamiltonian structures describing master equation and diffusion equation
systems are the subject
of ongoing exploration in stochastic processes research, where
the solution sets of $H=0$ near the deterministic subspace are
used to model quasistationary behaviors and rare transition events,
such as switching between states or noise-induced extinctions.
We have presented an application of 
these structures far away from the deterministic subsystem,
to approximate the
probability distribution of a process near an absorbing singular point,
where the WKB hypothesis does not hold and transient dynamics of
the limiting PDE
rather than its large-time limit behavior must be used to
identify the structure corresponding to the
quasistationary probability distribution of the finite-size system.

Quasistationary solutions in epidemic models can generally not be solved
exactly, so approximation techniques are crucial in analysis of
these processes.
We present an alternative approach to this approximation
problem, which may be extensible to other similar model settings
and whose full usefulness is yet to be discovered.
The WKB approximation and the Hamiltonian and
Lagrangian techniques
of analysis that it makes available are powerful and flexible,
and may have applications in subcritical disease settings that go
well beyond the quasistationary distribution.

Our exploration of cross-sectional prevalence data from trachoma trials, 
when the prevalence distributions are 
represented as curves on the Hamiltonian phase plane,
reveals a pattern of motion consistent with the motion on the
phase plane predicted by this analysis for a subcritical
transmission model. Thus it is consistent, at least qualitatively,
with a hypothesis that trachoma transmission in that trial setting
is in fact subcritical and stochastic.
This analysis fails to disconfirm that hypothesis,
though other explanations are possible.
In epidemiological settings where more data are available, it
may become possible to observe an upper limiting curve
in such a plot as well as the convergence to the vertical axis.
By revealing an emerging shape of the tail of the prevalence
distribution, information about that curve could contribute
to description of
the quasistationary behavior of the disease.
Such information also may contribute to
an estimate of its basic reproduction number, arrived at
independently of any estimate based on temporal change
in prevalences.

Beyond the one-variable birth-death models that we have analyzed,
the techniques that we explore here for study of quasistationary dynamics
may be of use with 
models with more stages of disease progression or differing transition rates,
multitype models,
models with patch or network structure (cf.\ \cite{hindes_epidemic2016}),
and other cases that are more complex than the simple models presented here.
In population biology, the SIS model we have discussed is also known
as a stochastic logistic model \cite{nasell1999},
and this analysis has promise for population
biology models that are similar but not identical to this
model.
While the primary goal in conservation biology is to preserve the
populations in question,
rather than to eradicate them as in epidemiology, declining populations
are clearly of interest and the models in use
may benefit from a similar analysis.
This analysis may be of use in other applications as well,
where quasistationary dynamics near an absorbing state
is of interest.

\section{Acknowledgments}

This study was supported by a Models of Infectious Disease
Agent Study (MIDAS) grant from the US NIH/NIGMS to the
University of California, San Francisco (U01GM087728),
by US NEI R01-EY025350,
and by a Research to Prevent Blindness Award.
IBS was supported by NRL base funding (N0001414WX00023)
and Office of Naval Research (N0001414WX20610).
We are grateful to two anonymous reviewers for helpful
comments on an earlier version of this manuscript.

\section{Conflict of Interest Statement}

The authors declare they each have no conflicts of interest.


\section{References}


\bibliographystyle{elsarticle-num}
\bibliography{subcritical,mainbib,trachoma,std}


\appendix

\section{Comparison to Poisson birth-death process}

The close approximation to the geometric by the SIS and other
transmission models when infective counts are small
can also be explained by comparing the transmission model to
a Poisson birth-death process,
that is, a process in which depletion
of the susceptible class is not accounted for \cite{bailey1975}.
The quasistationary limit of this process is the Yaglom
limit of its associated branching process
\cite{yaglom1947certain,harris_theory_1963}, and is exactly geometric.

Here we use the above Hamiltonian phase-plane analysis to
approximate the quasistationary limit of
a Poisson birth-death process with the same basic reproduction number
$R_0$, for comparison to the above results.

In the Poisson birth-death process,
the birth and death rates are the same as in the SIS model,
except for the absence of the nonlinear $S$ factor in the birth
rate:
\[ B(I) = R_0 I, \qquad D(I) = I. \]
It follows that the resulting Hamiltonian also is the same except for
the absence of that factor:
\begin{dmath*}
 H(x,p) = R_0 x \left( e^p -1 \right) + x \left( e^{-p} - 1 \right) \\
   = x \left( R_0 - e^{-p} \right) \left( e^p - 1 \right).
\end{dmath*}

As in the SIS case, the Hamiltonian factors into three parts,
corresponding to three intersecting components of the solution set
of $H=0$.
\begin{figure}
\centering
\includegraphics[width=0.5\textwidth]{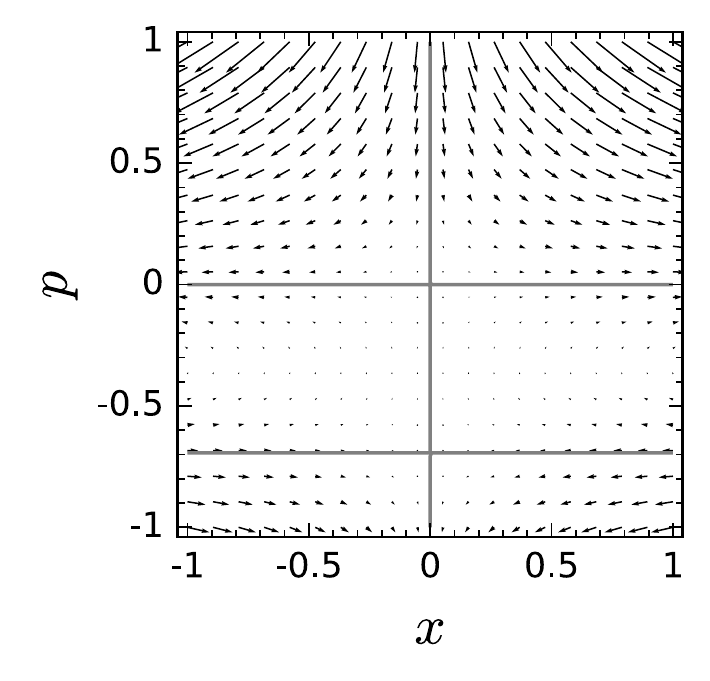}
\caption{ \label{fig:yaglom-plane-supercritical}
{\bf Phase plane for supercritical Poisson birth-death process} ($R_0=2$).
}
\end{figure}
\begin{figure}
\centering
\includegraphics[width=0.5\textwidth]{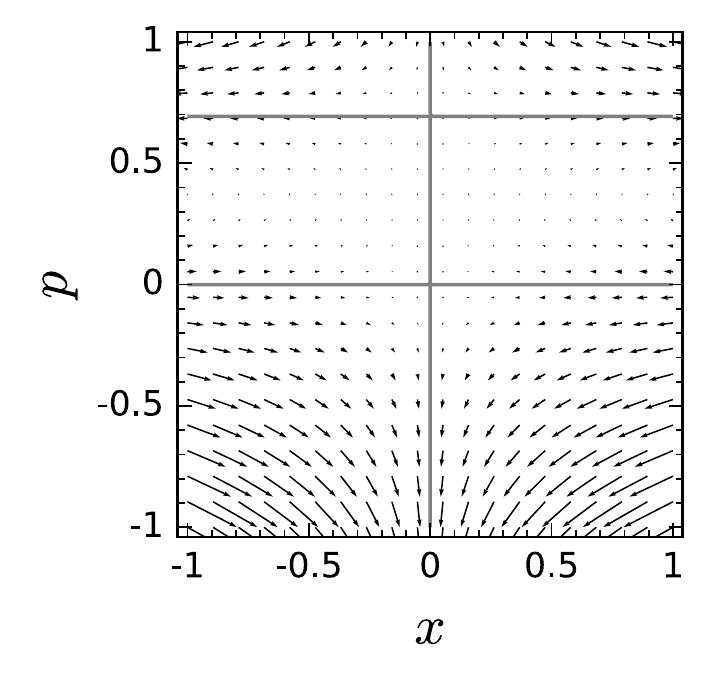}
\caption{ \label{fig:yaglom-plane-subcritical}
{\bf Phase plane for subcritical Poisson birth-death process} 
($R_0=1/2$).
}
\end{figure}
Again, two components are the vertical and horizontal axis.
The third, nontrivial solution
in this case is a horizontal line rather than a rising curve.
Unlike the SIS case, here the nontrivial curve does not intersect the
horizontal subsystem (except in the critical case, which we will leave
aside in this discussion).  

In the supercritical case (figure~\ref{fig:yaglom-plane-supercritical}),
the dynamics on the horizontal axis
(the deterministic subsystem) is similar to the SIS model in that
positive values rise away from zero, but with the difference that
they increase to infinity rather than to a finite endemic equilibrium
value.
In the subcritical case (figure~\ref{fig:yaglom-plane-subcritical}),
in which the birth-death process tends to
extinction, the behavior of the Hamiltonian system
in the $x\geq0$ half-plane is qualitatively
the same as for the subcritical SIS.
The only difference is in the form of the nontrivial curve that specifies
the quasistationary distribution.

The quasistationary curve is specified by the
equation $R_0=e^{-p}$, or $p=-\ln R_0$.
Substituting $\partial u/\partial x$ for $p$
produces the quasistationary solution for the action,
$u(x) = -x\ln R_0$,
and for the probability density,
$\phi(x) = C\,e^{-Nu(x)} = C\,(R_0)^I$.
This is equivalent to the geometric distribution with parameter $R_0$,
which is well known to be the quasistationary distribution of this
process.

We note that we can relate the geometric approximation to the geometry
of the Hamiltonian phase plane.
As discussed above, the geometric approximation to the quasistationary
distribution is characterized by the logarithmic slope
$\frac{\partial}{\partial I}\ln P(I,t)$.
The density function in terms of the fraction $x$ is
$\phi(x,t) = \Omega P(I,t)$, and the action $u$ is defined by
$\phi(x,t) = \Omega e^{-\Omega u(x,t)}$, or
$u(x,t) = -\ln (\phi(x,t)/\Omega)/\Omega$.
Combining,
\[ u(x,t) = -\ln P(I,t) / \Omega, \]
so that
\begin{dmath*}
  \frac{\partial}{\partial x} u(x,t)
    = \Omega \frac{\partial}{\partial I} (-\ln P(I,t) / \Omega) \\
    = - \frac{\partial}{\partial I}\ln P(I,t).
\end{dmath*}
But note that on the phase plane, the vertical coordinate $p$ is
identified with $\partial u/\partial x$;
so it follows that the parameter of the geometric distribution
approximating the quasistationary distribution near $I=0$
is revealed by the value of the nontrivial solution for $p$ at $x=0$,
that is, the
intercept where the limiting curve describing the quasistationary
distribution crosses the vertical axis.
That intercept is equal to $-\ln R$, where $R$ is the parameter
of the approximating geometric distribution.

We note that the horizontal line found in the Poisson birth-death model's
phase plane and the SIS process's limiting
curve converge on the same value $p=-\ln R_0$ at $x=0$, confirming visually
that this birth-death process is a good approximation for the SIS process
when its infective count is much smaller than $N$.

\end{document}